\documentclass[%
pacs,
superscriptaddress,
twocolumn,
showpacs,
 amsmath,amssymb,
 aps,
prl,
footinbib,
]{revtex4-1}

\usepackage{amsmath,graphicx}
\usepackage{pstricks}
\usepackage{psfragx}
\usepackage{verbatim}
\usepackage{color}

\usepackage{graphicx}
\usepackage{dcolumn}
\usepackage{bm}
\usepackage{version}

\def\eqd{\,{\buildrel d \over =}\,}

\begin{document}


\title{Anomalous fluctuations of currents in Sinai-type random chains with strongly correlated disorder}

\author{Gleb Oshanin}
\email{oshanin@lptmc.jussieu.fr}
\affiliation{Laboratoire de Physique Th{\'e}orique de la Mati{\`e}re
Condens{\'e}e (UMR CNRS 7600), Universit{\'e} Pierre et Marie Curie (Paris 6) -
4 Place Jussieu, 75252 Paris, France}
\author{Alberto Rosso}
\email{alberto.rosso@lptms.u-psud.fr}
\affiliation{Laboratoire de Physique Th{\'e}orique et Mod{\`e}les  Statistiques (UMR CNRS 8626),
    Universit\'e de Paris-Sud,
Orsay Cedex, France}
\author{Gr\'egory Schehr}
\email{gregory.schehr@lptms.u-psud.fr}
\affiliation{Laboratoire Physique Th{\'e}orique et Mod{\`e}les  Statistiques (UMR CNRS 8626),
    Universit\'e de Paris-Sud,
Orsay Cedex, France}

\date{\today}

\begin{abstract}
We study properties of a random walk in a generalized Sinai model (SM), in which a quenched random potential
is
 a trajectory of a \textit{fractional} Brownian motion with arbitrary Hurst parameter $H$, $0< H <1$, so that the random force field displays strong spatial correlations. In this case, the disorder-average mean-square displacement (MSD) grows in proportion to $\log^{2/H} (n)$, $n$ being time. We
 prove that moments of arbitrary order $k$
 of the steady-state current $J_L$ through a finite segment of length $L$ of such
  a chain decay as $L^{-(1-H)}$, independently of $k$, which suggests that despite a logarithmic confinement the average current is much higher than its Fickian counterpart in homogeneous systems. Our results reveal a paradoxical behavior such that, for fixed $n$ and $L$, the MSD \textit{decreases} when one varies $H$ from $0$ to $1$, while the average current \textit{increases}. This counter-intuitive behavior is explained via an analysis of representative realizations of disorder.
\end{abstract}

\pacs{05.40.-a, 02.50.-r,05.10.Ln}

\maketitle

Since the pioneering works \cite{kesten,derrida,sinai}, random walks (RWs) in random media
attracted a considerable attention. In part,
due to a general interest in
dynamics in disordered systems,
but also because such RWs found many physical applications, including
dynamics of the helix/coil phases boundary
in a random heteropolymer \cite{pgg,redner},
a random-field Ising model
 \cite{bruinsma,nattermann}, 
dislocations in disordered crystals \cite{harth}, mechanical unzipping
of DNA \cite{walter},  translocation of biomolecules
through nanopores \cite{9} and  molecular
motors \cite{10}. Some functionals arising here, e.g., probability currents in finite samples,
 show up in mathematical finance~\cite{alain,greg}. 
 Other examples can be found in~\cite{comtet,hughes,sheinman}.

In the discrete formulation,
a RW evolves in a discrete time on a lattice. 
At each time step the walker
jumps from site $X$ to either 
site $X + 1$ with the site-dependent probability $p_X = \frac{1}{2} (1 + \varepsilon \cdot s_X)$,
or to the site $X - 1$ with the probability
$q_X = 1-p_X$, where the amplitude $0 < \varepsilon < 1$ measures the strength of the disorder and
$s_X$ are quenched, independent and identically distributed (i.i.d.) random variables (r.v.). One often  assumes  binomial r.v., i.e., $s_X = \pm 1$ with probabilities $p$ and $1-p$, respectively. 



In case of no global bias ($p=1/2$), i.e. for the so-called Sinai model (SM), a remarkable result \cite{sinai} is that
for a given environment $\{p_X\}$
the squared displacement
\begin{equation}
\label{sinai}
X_n^2 \sim m(\{p_X\}) \, \ln^4(n) \,,
\end{equation}
as $n \to \infty$ with probability almost $1$, where
$m(\{p_X\})$ is a function of the environment only~\cite{note2}. Another intriguing feature
of the SM concerns transport properties. It was revealed by analysing
the probability current $J_L$ through a finite Sinai chain of length $L$, that
the disorder-average current decays as $1/\sqrt{L}$ \cite{gleb,gleb1,gleb2,cecile}.
Curiously enough,
despite a logarithmic confinement (\ref{sinai}), the disorder-average current
appears
to be anomalously high, so that
such disordered chains offer on average less resistance with respect to transport
of particles than homogeneous chains (all $p_X \equiv 1/2$)
for which one finds Fick's law $J_L  \sim 1/L$. In absence of disorder, deviations from Fick's law can also be found for
L\'evy walks~\cite{dhar_derrida}. Full statistics of the current has been recently computed for ASEP model \cite{kirone}.


It is well-known that a RW in uncorrelated random environment $\{p_X\}$ can be
considered as a one in presence of a random potential $V_L$, which represents
itself a RW in space.
Indeed, on scales $L$ a RW "explores" the potential
%
\begin{equation}
\label{walk}
V_L = \sum_{X = 1}^{L - 1} \ln\left(\frac{p_X}{q_X}\right) = \sigma \sum_{X = 0}^{L - 1} s_X \,, \sigma = \ln\left(\frac{1 + \varepsilon}{1 - \varepsilon}\right)  \,,
\end{equation}
which is just a RW trajectory
with step length $\sigma$. Standard SM, in which the $s_X$'s are uncorrelated
is now well understood. On the contrary, there hardly exist analytical results for 
the case where these r.v. are strongly correlated. Such correlations are important, e.g., for the dynamics of the helix-coil boundary
in random heteropolymers, where the 
chemical units are usually strongly correlated \cite{gros3}. 
They are also currently studied 
in mathematical finance, improving 
the standard Black-Sholes-Merton (BSM) model~\cite{fBm_finance}. 
Any exact result for such situations would thus be welcome.

In this Letter, we study properties of random walks in random environments
 in which the transition probabilities $\{p_X\}$ are strongly correlated
  so that the potential $V_L$ in~(\ref{walk}) is a fractional Brownian motion (fBm):
$V_L$ is Gaussian, with $V_{L=0}=0$ and moments
\begin{equation}
\label{def}
\mathbb{E}\left\{ V_L \right\} = 0 \,, \,\, \mathbb{E}\left\{ (V_L - V_{L'})^2 \right\} = \sigma^{2} |L - L'|^{2 H}\,,
\end{equation}
where $\mathbb{E}\{\ldots\}$ here and henceforth denotes averaging over realizations of $V_L$ and $0 < H < 1$. The case $H = 1/2$ corresponds to the original SM. For $H < 1/2$ the potential is subdiffusive while for $H > 1/2 $ it is superdiffusive.

The mean squared displacement $\mathbb{E}\{X_n^2\}$ in a correlated random environment can be estimated as follows. Assuming Arrhenius' law for the activated dynamics~\cite{comtet},
the time $n_L$ required for a particle to diffuse in a disordered potential $V_L$
 over a scale $L$, is of order $n_L \sim e^{V_L^*}$, where $V_L^*$ is a typical
 energy barrier. For $V_L$ in (\ref{def}), $V_L^* \sim \sigma L^H$, so that
for sufficiently large times $n$
\begin{eqnarray}
\label{1}
\mathbb{E}\left\{X_n^2\right\} \sim \sigma^{-2/H} \ln^{2/H}(n) \,.
\end{eqnarray}
Our focal interest here is in understanding the behavior
of the disorder-average
current $J_L$ through a finite sample (of length $L$)
of such a disordered chain, of its moments of arbitrary order,
and eventually, of the full probability density function (pdf) of $J_L$. We proceed to show
that, while its typical value is exponentially small $J_{{\rm typ}} \sim \exp{(-L^H)}$, all its moments decay algebraically
\begin{equation}
\label{2}
\mu_k(L) \equiv \mathbb{E}\left\{ \left({J_L}\right)^k \right\} \sim A_k {L}^{-\theta}\;, \; L \gg 1 \,,
\end{equation}
where $\theta = 1 - H$ is the persistence exponent of
the fBm \cite{krug,molchan}. Recall that the persistence exponent associated to a stochastic process characterizes the algebraic decay of its survival probability $S(n)\sim n^{-\theta}$~\cite{majumdar,aurzada_simon}. The $L$-independent constants $A_k$ depend in general, on the microscopic details (as the lattice discretization). 

The result in (\ref{2}) is rather astonishing: a) it states that $\mu_k(L)$ for arbitrary  order $k$ 
decay in the same way. b) for arbitrary $H$, $0 < H < 1$,
the disorder-average current in such random chains is larger than the Fickian current in homogeneous systems and c) on comparing
(\ref{1}) and (\ref{2}) for fixed $n$ and $L$ sufficiently large, and varying $H$, one concludes that $\mathbb{E}\{X_n^2\}$ \textit{increases} when $H$ goes from $1$ to $0$, while the disorder-average current \textit{decreases}, which is an absolutely counter-intuitive and surprising behavior.

In what follows we prove (\ref{2}) and explain
this astonishing behavior
using three complementary approaches: (i) a rigorous one, based on exact bounds, for the discrete RW in a fBm potential, (ii) scaling arguments for the continuous-space and -time version, which also allows to study the whole pdf of $J_L$ and (iii) via numerical simulations. We argue that (\ref{2}) holds for any potential $V_L$, which is the trajectory of a stochastic process with persistence exponent $\theta$: as a matter of fact, such a behavior of $\mu_k(L)$ is
dominated by the configurations of $V_L$ which drift to $-\infty$ without re-crossing the origin and
 occur with a probability $\sim L^{-\theta}$, yielding the $L$-dependence in~(\ref{2}).

\begin{figure}[ht]
\includegraphics[width=\linewidth]{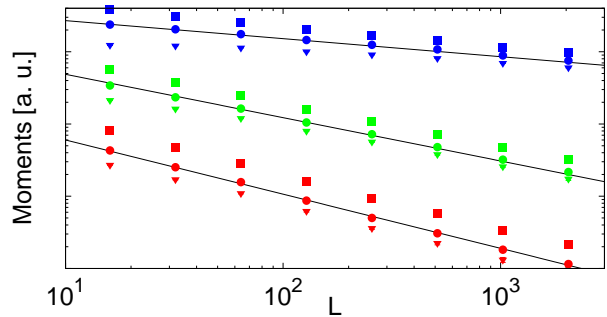}
\caption{(Color online) $\mathbb{E}\left\{ J_L \right\} $  (squares) and $\mu_k(L)$ with $k=2$ (circles) and $k=3$ (triangles) vs $L$
for the fBm with   $H =0.75$, $0.4$ and $0.25$ (from top to bottom).
The solid line is $L^{-\theta}$ (\ref{2}) with $\theta=1-H$. The temperature $T=0.25$ and averaging is performed over $10^5$ samples.
We use arbitrary units [a.u.] because we vertically shift the data (by a factor $20$ for $H=3/4$, $5$ for $H=0.4$ and $1$ for $H=0.25$). In any case, the prefactors $A_k$ are non-universal and model dependent.}
\label{fig_moments}
\end{figure}

%
%
%
%

Consider first the discrete chain, take a finite segment of length $L$
and impose
 fixed concentrations of particles at the endpoints,
 $P_0$ and $P_L$. For a fixed environment  $\{p_X\}$, the steady-state current is given by
\cite{gleb,gleb1}
\begin{eqnarray}
\label{current}
J_L = \frac{D_0 P_0}{\tau_L} - \frac{D_0 P_L}{\tau^*_L},
\end{eqnarray}
where $D_0 = 1/2$ is the diffusion coefficient of a homogeneous chain, $\tau_L$ is the so-called Kesten variable \cite{kesten2}:
\begin{eqnarray}
\label{tau}
\tau_L = 1 + \frac{p_1 }{q_1} + \frac{p_1 p_2}{q_1 q_2} + \ldots + \frac{p_1 p_2 \ldots p_{L - 1}}{q_1 q_2 \ldots q_{L - 1}} \,,
\end{eqnarray}
and $\tau_L^*$ is obtained from (\ref{tau}) by replacements $p_k \to q_{L-k}$ and $q_k \to p_{L-k}$.
%
%
Thinking of $L$ as "time", one notices that $\tau_L$ and $\tau^*_L$
are time-averaged discretized geometric fractional Brownian motions (they can
be thought of as the "prices" of Asian options
within the framework of the fractional BSM model \cite{greg}).
 Note that in absence of a global bias
$\mathbb{E}\{ 1/\tau_L\} = \mathbb{E}\{ 1/\tau^*_L\}$,
and hence, without any loss of generality we set $P_L = 0$ in what follows. Thus, combining (\ref{walk}, \ref{current}, \ref{tau}) and setting $P_0=1$ yields
\begin{eqnarray}\label{explicit_J}
J_L = \frac{1}{2}  \left(1 + \sum_{l=1}^{L-1} \exp\left(V_l\right)\right)^{-1} \;.
\end{eqnarray}
For typical realizations
of
$\{p_X\}$, the size of $|V_l|$ is $\mathcal{O}(l^H)$ so that the typical current $J_{{\rm typ}} $ is $J_{{\rm typ}} \sim \exp\left(- L^H\right)$.


To obtain an upper bound on $\mu_k(L)$, consider a given realization of the sequence $V_1,V_2, \ldots,V_{L-1}$ and denote the maximal among them as $V_{\max} = {\max}_{0\leq i \leq L-1}V_i$. From~(\ref{tau}) one has $\tau_L  = (1 + \sum_{l=1}^{L-1} V_l)\geq \exp\left(V_{\max}\right)$, so that
$J_L^k \leq \left( {1}/{2}\right)^k \exp\left( - k V_{\max}\right)$.
Since $\exp\left( - k V_{\max}\right) \to 0$ as $L \to \infty$ (recall that $V_{\max} \sim L^H$) the average value of $\exp\left( - k V_{\max}\right)$ is dominated by configurations with $V_{\max} \to 0$. The asymptotic behavior of the pdf  $P_L(V_{\rm max})$ for fixed $V_{\max}$ and large $L$ is known~\cite{krug,molchan}, yielding $\ln P_L(V_{\max}) = \theta \ln L^{-1} + {\cal O}(1)$, where $\theta = 1 - H$ is the persistence exponent~\cite{note}. Hence, we have
\begin{equation}
\label{upper_2}
\mu_k(L)
\leq {B_k}{L^{H - 1}} \,, L \gg 1\,,
\end{equation}
where $B_k$ is an $L$-independent constant.

To determine a lower bound on $\mu_k(L)$
we follow~\cite{gleb,gleb1,cec} and make the
following observation: averaging ~(\ref{explicit_J}) is to be performed over the entire
 set $\Omega$ of all possible trajectories $\{V_l\}_{1 \leq l \leq L}$. Since $\tau_L > 0$, a lower bound on
$\mu_k(L)$
can be straightforwardly obtained if one averages instead  over some
finite subset $\Omega' \subset \Omega$ 
of trajectories with some prescribed properties, that is
%
$\mu_k(L) \geq \mathbb{E}_{\Omega'}\left\{ J_L^k\right\}$.
We choose $\Omega'$ as the set
comprising
all
possible trajectories $\{V_l\}_{0 \leq l \leq L}$ which, starting at the origin at $l = 0$,
never cross the deterministic curve
$Y_l = Y_0 - \alpha \ln(1+l)$ with $Y_0 > 0$ and $\alpha > 1$. For any such trajectory
$\tau_L = 1 + \sum_{l = 1}^{L - 1} \exp\left(V_l\right)$ is bounded from above by $\sum_{l = 0}^{L - 1} \exp\left(Y_l\right)$, which, in turn, is bounded from above by $\exp(Y_0) \zeta(\alpha)$,
where $\zeta(\alpha)$ is the zeta-function.  Hence, we have
$\mu_k(L)  \geq (\exp(Y_0) \zeta(\alpha)/2)^{-k} \, \mathbb{E}_{\Omega'}\left\{1\right\}$,
where $\mathbb{E}_{\Omega'}\left\{1\right\}$ is, by definition, the survival probability, $S_L$ up to time $L$, for a fBm, starting at the origin in presence of a "moving trap" evolving via $Y_l = Y_0 - \alpha \ln(1 + l)$. 

For standard Brownian motion ($H=1/2$)  in presence of a trap which moves as $- l^z$, the leading large-$L$ behavior of
the survival probability $S_L$
is exactly the same as in the case of an immobile trap, provided that $z < 1/2$~\cite{krap}. It is thus physically plausible to suppose that the same behavior holds for a more general Gaussian process such as a fBm. That is,
one expects that for any $H > 0$ the leading
large-$L$ behavior of $\mathbb{E}_{\Omega'}\left\{1\right\}$ will be exactly the same
for an immobile trap and for
a logarithmically moving trap, i.e., that $S_L = \mathbb{E}_{\Omega'}\left\{1\right\} \sim Y_0^{\theta/H}/L^{\theta}$ as $L \to \infty$ \cite{krug,molchan}, where $\theta = 1 - H$. In fact, this can be shown rigorously \cite{aurzada_fbm,aurz}. Consequently, we find
\begin{equation}
\label{lower}
\mu_k(L) \geq {D_k}{L^{H - 1}} \;, L \gg 1\,,
\end{equation}
where $D_k$ is independent of $L$. Note that the bounds in~(\ref{upper_2}) and (\ref{lower}) show
 the same $L$-dependence and thus yields the exact result announced in (\ref{2}).

We now turn to 
a continuous-time and -space dynamics in a disordered fBm potential. The position $x(t) \in [0,L]$ of a particle at time $t$ obeys a Langevin equation :
$\dot x = - V'(x) + \eta(t)$,
where $V'(x)$ is a quenched random force  such that $V(x)$ is a fBm with Hurst exponent $H$~(\ref{def}) and $\eta(t)$ is a Gaussian thermal noise of zero mean and covariance $\langle \eta(t) \eta(t')\rangle = 2 T \delta(t-t')$. The steady-state current and the  concentration profile $C(x)$ can be obtained from the corresponding Fokker-Planck equation 
\begin{eqnarray}\label{start_continuum}
&&J_L = {T}\left({\int_0^L \, \exp{[{V(x)}/{T}]}\, dx}\right)^{-1} \;, \\
&&C(x) = \frac{J_L}{T} \int_x^L dx' \, \exp{[(V(x')-V(x))/T]} \;, \nonumber
\end{eqnarray}
[see (\ref{current}, {\ref{explicit_J}}) with $D_0=T$ and $P_0=1$] \cite{remark}. The total number of
particles is then $N_L = \int_0^L C(x) \, dx$. 
We focus next on the moments and on the pdf of $J_L$ (\ref{start_continuum}).

\begin{figure}
\centering
\includegraphics[width=\linewidth]{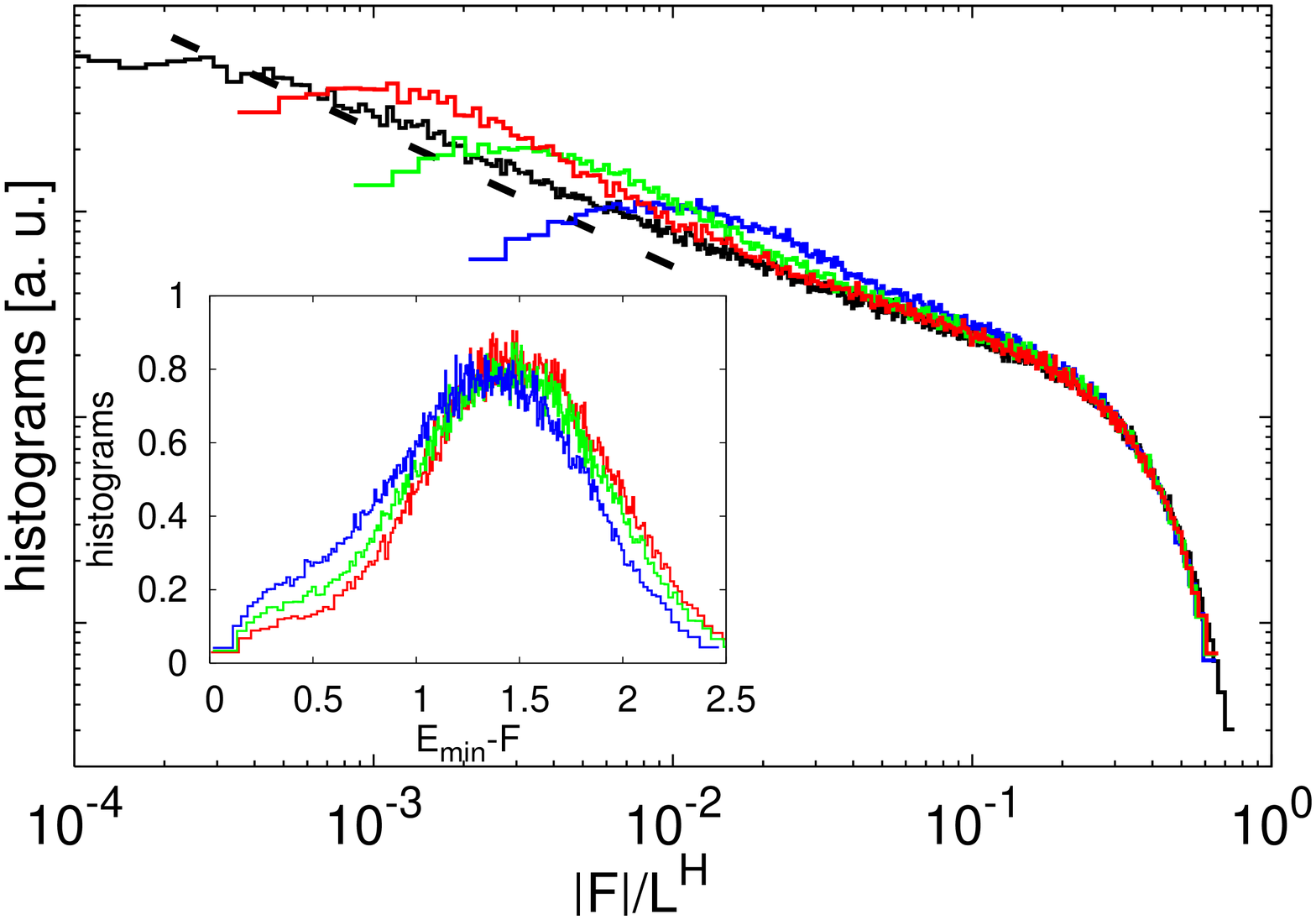}
\caption{(Color online) Pdf of the free energy $-F/L^H$ for different system sizes: $L=64 \,(\rm blue), 256 \,(\rm green), 1024 \, (\rm red)$. The black curve represents the distribution of $E_{\rm min}/L^H$ for $L=4096$. {\bf Inset:} Pdf of $E_{\min} -F$ for different system sizes: $L=64 \,(\rm blue), 256 \,(\rm green), 1024 \, (\rm red)$. Histograms are computed using $10^5$ samples and setting $T=1$ and $H=3/4$. The dashed line corresponds to $(|F|/L^H)^{-2/3}$~(\ref{eq:dist_f}).}\label{fig2}
\end{figure}

Instead of $J_L/T$, which can be viewed as the inverse of the 
SM partition function, we study the pdf $\Pi_{T=0}(F)$ of
the free energy $F = T \log(J_L/T)$.
Consider first $T \equiv 0$, in which case $F=E_{\min} = \min_{0\leq x \leq L} V(x)$. Recalling
that $V(0) = 0$ ($E_{\min} < 0$), 
the cumulative distribution $q_L(E) = {\Pr}(E_{\min} > - E)$, (with $E>0$), coincides with the probability that up to 'time' $L$, $V(x)$ starting at $E$ at $x=0$ 'survives'
in presence of an absorbing boundary at $V=0$. For self-affine process, $q_L(E)$ takes the scaling form $q_L(E) = Q(E/L^H)$: for $L \gg E^{1/H}$, $q_L(E)$ behaves algebraically \cite{majumdar},
$q_{L}(E) \sim {E^{\theta/H}}/{L^\theta}$
($\theta = 1 - H$ for fBm~\cite{krug,molchan}), while for $L \ll E^{1/H}$, $q_{L}(E)$ is of order one. Hence, one has for $\Pi_{T = 0}(F) = \partial_E q_L(E)\big|_{E = -F}$
\begin{eqnarray}\label{eq:dist_f}
\Pi_{T=0}(F) =
\begin{cases}
& 0 \;, \; F > 0\\
& L^{-\theta} |F|^{\theta/H-1} \;, \; - L^H \ll F < 0 \\
& \exp(-F^2/2 L^{2H}) \;, \; F \ll - L^H \;.
\end{cases}
\end{eqnarray}
Lastly, the regime $F \ll - L^H$ corresponds to a fraction of paths $V(x)$ that propagate from $E$ to zero in a 'time' $L$. In general, the tail
of this probability coincides with the one of the free propagator, which is Gaussian for fBm. What happens at finite $T$ where $F = E - T S$ is now the balance between the energy $E$ and the entropy $S$~? One expects a particle to be localized close to the minimum $E_{\min}$, which is of order ${\cal O}(L^H)$, while the maximum entropy $\sim {\cal O}(\ln L)$. Hence when $L \gg 1$ the main contribution to $F$ comes from $E_{\min}$ so that for a given sample at finite $T$, $F$ will be very close to $E_{\min}$. This is corroborated by numerical simulations (see Fig.~\ref{fig2}).

We now come back to the current distribution. Very small currents,  $J_L \ll J_{\rm typ} \sim \exp(-L^H)$, correspond to $F \ll - L^H$ in~(\ref{eq:dist_f}) and one obtains that $P(J_L)$ is log-normal, $\ln P(J_L) \propto {-{\ln^2(J_L/T)}}$. Within the opposite limit, $J_L \gg J_{\rm typ}$, one finds from~(\ref{eq:dist_f}) that
\begin{eqnarray}\label{eq:inter_regime}
P(J_L) \sim \frac{[\log(J_L/T)]^{\theta/H-1}}{J_L L^\theta} \;.
\end{eqnarray}
This power-law behavior holds up to a large cut-off value $J_{\max}$. At $T=0$ we have a sharp cut-off at $J_{\max}=1$ (\ref{eq:dist_f}), while at a finite $T$, $P(J_L)$ has a fast decay which depends on the fluctuations of $V(x)$ at a short length scale close to the origin $x=0$.
For $L \gg 1$, $\mu_k(L)$ are dominated by the regime where $J_{\rm typ} \ll J_L < J_{\max} \sim {\cal O}(1)$ (\ref{eq:inter_regime}), such that one gets
$\mu_k(L) \sim 1/{L^\theta}$ (\ref{2}). This calculation shows that (rare) negative persistent potential leads to very large currents. We observe that these rare persistent profiles also exhibit large barriers, growing like $L^H$. These barriers stop the particle diffusion and are responsible for the subdiffusive behavior of the mean square displacement. One could expect that these barriers should also affect the behavior of the current. However, by looking at the steady state concentration profile $C(x)$ (\ref{start_continuum}), one can see that large barriers induce a very large number of particles in the system located in the deep valleys of the potential $V(x)$, which allows to sustain a large current.

In our numerical simulations we consider a discrete random potential $V_k$, $k=0,1,\ldots L-1$, with $\sigma^2=1$, which displays fBm correlations (\ref{def}). We use a powerful algorithm~\cite{rosso_fbm,GRS10}, which allows to generate very long samples of fBm paths. For each sample, we compute the current, $J_L = T [\sum_{k=0}^{L-1} \exp(-V_k/T)]^{-1}$, the free energy $F = T \ln (J_L/T)$ and the ground state energy $E_{\min} = \min_{k} V_k$. In Fig.~\ref{fig_moments} we plot the first three moments as a function of $L$ for different values of $H$. These plots show a very good agreement with our analytical predictions in~(\ref{2}). In Fig.~\ref{fig2}, we show that the pdf of the rescaled free energy, $F/L^H$, at finite temperature $T$ converges to the pdf of the rescaled ground state energy $E_{\min}/L^H$. The reason for this is that, for each sample, the difference between $F$ and $E_{\min}$ grows very slowly with $L$, probably logarithmically (inset of Fig. \ref{fig2}). In the rescaled variables, this difference vanishes when $L \to \infty$.



\begin{figure}[ht]
\includegraphics[width=0.8\linewidth]{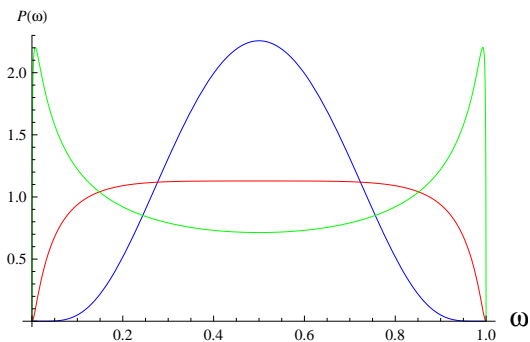}
\caption{\label{fig} (Color online) The pdf in Eq.~(\ref{omega}) for $\sigma^{2} L^{2 H} = 1/2$ (blue), $\sigma^2 L^{2 H} = 2$ (red)  and $\sigma^2 L^{2 H} = 5$ (green).}
\end{figure}

We close with an observation
that such chains show a transition to a diode-like behavior, when  $\xi = \sigma^2 L^{2 H}$ exceeds some critical value $\xi_c$. Consider a chain in which at site $X = 0$ we maintain a fixed concentration $P_0 = 1$ of, say, "white" particles and place a sink for them at $X = L$. At $X = L$ we introduce a source which maintains concentration $1$ of "black" particles, and place a sink for them at site $X = 0$. The particles are mutually noninteracting. For a fixed $\{p_X\}$ we have counter-currents of white ($J_L^w$) and black ($J_L^b$) particles, which obviously obey, on average, $\mathbb{E}\left\{ (J_L^w)^k\right\} \equiv \mathbb{E}\left\{ (J_L^b)^k\right\}$ for any $k>0$. 

Consider next the random variable: 
$\omega = {J_L^w}/{(J_L^w + J_L^b)} = {\tau_L^*}/{(\tau_L^* + \tau_L)}$,
which probes the likelihood of an event that for a fixed $\{p_X\}$ one has $J_L^w = J_L^b$. The pdf of $\omega$ can be calculated exactly to give
\begin{eqnarray}
\label{omega}
P(\omega) = \frac{1}{\sqrt{2 \pi} \omega (1-\omega) \sigma L^H} \exp\left(- \frac{\ln^2\left(\frac{1 - \omega}{\omega}\right)}{2 \sigma^{2 } L^{2 H}}\right)\,.
\end{eqnarray}
%
Remarkably,  $P(\omega)$ in (\ref{omega}) changes the modality when $\xi$, (which defines the value of a typical barrier), exceeds a critical value $\xi_c =2$ (see Fig.~\ref{fig}). For short chains (or small $\sigma$) $P(\omega)$ 
is unimodal and centered at $\omega = 1/2$: any given sample
is transmitting particles in both directions equally well and, most probably, $J_L^w = J_L^b$. For $\xi = \xi_c$
the pdf is nearly uniform (except for narrow regions at the edges) so that \textit{any} relation between $J_L^w$ and $J_L^b$
is equally probable. Finally, for $\xi > \xi_c$ (sufficiently strong disorder and/or a long chain) the symmetry is broken and $P(\omega)$ becomes bimodal with a local minimum at $\omega = 1/2$ and two maxima close to $0$ and $1$. This means that a given sample is most likely permeable only in one direction.

\acknowledgments

We thank G. Biroli for a useful discussion. GO is
partially supported by the ESF Research Network "Exploring the Physics
of Small Devices",
AR - by ANR grant 09-BLAN-0097-02 and GS - by ANR grant 2011-BS04-
013-01 WALKMAT. This project was partially supported by the Indo-French Centre for the Promotion of
Advanced Research under Project 4604-3.


\begin{thebibliography}{99}
\bibitem{note} As a matter of fact, this result
can be made more precise by writing the distribution $P_L(V_{\max})$ in a scaled
form $P_L(V_{\max}) = L^{-H} f(y = V_{\max}/L^H)$. This implies that actually
$P_L(V_{\max}) \sim V_{\max}^{(1-2H)/H}/L^{1 - H}$ as $L \to \infty$.
\bibitem{remark} We note that the functional in (\ref{start_continuum}) was studied within a different context in
\cite{molchan,MRZ09}, and it was shown that its first moment decreases as $L^{-\theta}$ with $\theta = 1 - H$.
See also different exact results \cite{gleb2,cecile}
for the $H = 1/2$ case.
\end{thebibliography}
\end{document}